\documentstyle[sprocl,epsf]{article}

\def\Journal#1#2#3#4{{#1} {\bf #2}, #3 (#4)}

\def\PRL{\em Phys. Rev. Lett.}
\def\PRD{{\em Phys. Rev.} D}

\def\beq{\begin{equation}}
\def\eeq{\end{equation}}
\def\bea{\begin{eqnarray}}
\def\eea{\end{eqnarray}}
\newcommand\bzeta{\mbox{\boldmath$\zeta$}}
\newcommand\bp{{\bf k}}

\newcommand\E{\epsilon}
\newcommand{\ds}[1]{#1 \hspace{-0.5em}/}  

\begin{document}

\title{
\hfill {\normalsize KUNS-1656 }\\
Ferromagnetism of quark liquid and magnetars}

\author{Toshitaka Tatsumi}

\address{Department of Physics, Kyoto University, 
Kyoto 606-8502, Japan\\E-mail: tatsumi@ruby.scphys.kyoto-u.ac.jp} 


\maketitle\abstracts{ Spontaneous magnetization of quark liquid is
examined on the analogy with that in electron gas. 
It is pointed out that quark liquid has potential to be 
ferromagnetic at rather low densities, around nuclear saturation density. 
Some comments are given as for
implications on magnetars. 
}

\section{Introduction}

Recently a new type of neutron stars with extraordinary magnetic
field, usually called magnetars,  
has attracted much attention in connection with pulsars associated
with soft-gamma-ray
repeaters (SGR) and anomalous X-ray pulsars (AXP). There have been
known several
magnetar candidates such as SGR 1806-20 and
SGR 1900+14 \cite{ko}. Various analysis including the
$P-\dot P$ curve have indicated an intense 
magnetic field of $O(10^{14 - 15})$ G, while 
ordinary radio pulsars have a magnetic field of 
$O(10^{12 - 13})$G.

The origin of the strong magnetic field in compact stars, especially 
neutron stars, has been an open
problem. Recent discovery of magnetars seems to renew this
problem. Conservation of the magnetic flux during the collapse of a
main sequence star has been a naive idea to understand the magnetic field
in neutron stars \cite{cha}. Then the strength $B$ should be 
proportional to $R^{-2}$, where $R$ is a radius of a star; 
for example, the sun, 
a typical main sequence star, 
has a magnetic field of $O(10^3)$G with the radius $R\sim
10^{10-11}$cm. By decreasing the radius to $10^6$cm for neutron stars
$B=O(10^{12})$G, which is consistent with observations for
radio pulsars. However, if this argument is extrapolated to explain the
magnetic field for magnetars, we are lead to a contradiction:
their radius should be 
$O(10^4)$cm to get an increase in $B$ by a factor of $\sim 10^{12}$, 
which is much less than the Schwartzschild radius of 
neutron stars with the canonical mass $M=1.4M_\odot$,
$R_{Sch}=2GM/c^2=4\times 10^5$cm.

When we compare the energy scales for systems such as 
atomic system ($e^-$), nucleon system ($p$) and quark system ($q$), 
we can get a hint
about the origin of the magnetic field.  In Table 1. we list the
interaction energy, $E_{int}=\mu_iB$, of the magnetic field  
$B=10^{15}$G and each constituent with the Dirac magnetic moment, 
$\mu_i=e_i\hbar/2m_ic$. We also list a typical energy scale $E_{typ}$ 
for each system. 
\begin{table}
\begin{center}
\begin{tabular}{cccc}
 &$e^-$ & p & q\\ \hline\hline
$m_i$[{\rm MeV}]& 0.5 & $10^3$ & 1- 100\\
$E_{int}$[{\rm MeV}]~~~ & 5 - 6  &~~~$2.5\times 10^{-3}$~~~& 
$2.5\times 10^{-2}$ - 2.5\\ \hline
$E_{typ}$ & $O({\rm KeV})$ & $\geq O({\rm MeV})$ & $\geq O({\rm MeV})$
\end{tabular}
\end{center}
\caption{}
\end{table}
Then we can see that $E_{typ}\ll E_{int}$ for the electron system,
while $E_{typ}>E_{int}$ for the nucleon and the quark systems; that is,  
the strength of $O(10^{15})$G is
very large for the former system with the electromagnetic interaction, while it
is not large for the latter systems with the strong interaction. Hence it
may be conceivable that the strong interaction should easily produce
the magnetic field of the above magnitude.
Since there is a bulk hadronic matter 
beyond nuclear saturation 
density ($n_0\sim 0.16$fm$^{-3}$) inside neutron stars, 
it should be interesting to consider the hadronic 
origin of the magnetic field;  
ferromagnetism or spin-polarization of
hadronic matter may give such magnetic field. 

In 70's, just after the first discovery of pulsars, there have been
done many works about the possibility of the ferromagnetic transition
in dense neutron matter, using $G-$matrix calculations or variational
calculations with the realistic nuclear forces. Through these works there
seems to be a consensus that ferromagnetic
phase, if it exists, should be at very high densities, and there is no
transition at rather low densities relevant to neutron stars \cite{pan}.

We consider here
the possibility of ferromagnetism of quark liquid interacting with the 
one-gluon-exchange (OGE) interaction \cite{ta}.
One believes that there are deconfinement transition and  
chiral symmetry restoration at finite baryon density, 
while their critical
densities have not been fixed yet. One interesting suggestion is that
three-flavor symmetric quark matter (strange matter) 
may be the true ground state of
QCD at finite baryon density \cite{chi,far}. If this is the case, 
quark stars (strange quark stars), can exist in a different branch
from the neutron-star branch in the mass-radius plane \cite{mad}. 
Usually one implicitly assumes that the ground state of quark matter
is unpolarized. We examine here the possibility of polarization of
quark matter.
We shall see our results should
give an origin of the strong magnetic field for magnetars 
in the context of strange quark-star scenario.

\section{Spontaneous magnetization of quark liquid}
\subsection{Relativistic formulation}
Quark liquid should be totally color singlet (neutral), which means
that only the exchange interaction between quarks is relevant there.
This may remind us of electron system with the Coulomb force
in a neutralizing positive charge background.
In 1929 Bloch first suggested a possibility of ferromagnetism of
electron system \cite{bl}. He has shown that there is a trade off between the
kinetic and the exchange energies as a function of a polarization parameter, 
the latter of which favors the spin 
alignment due to a quantum effect; electrons with the same spin
orientation can effectively avoid the Coulomb repulsion due to the
Pauli exclusion principle. When the energy gain due to the spin
alignment dominate over the increase in the kinetic energy at some
density, the unpolarized electron gas suddenly turns into the
completely polarized state.
   
In the following we discuss the
possibility of ferromagnetism of quark liquid on the analogy with
electron gas (Fig.~1). 

\begin{figure}[h]
\centerline{
\epsfysize=4cm
\epsfbox{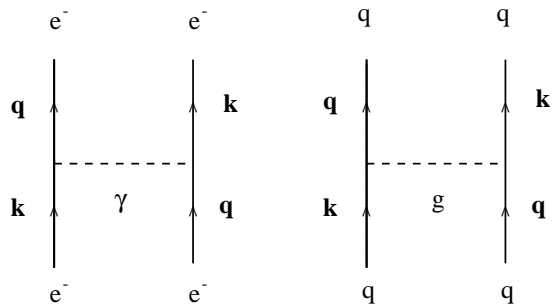}
}
\caption{Exchange interactions for electrons with the Coulomb force 
(left) and quarks with OGE interaction (right).}
\end{figure}
It is to be noted that there is one big difference
between them; quarks should be treated in a relativistic way. The
concept of the spin orientation is not well defined in relativistic
theories, while each quark has two polarization degrees of
freedom. Here we define the spin-up and -down states in the rest frame of
each quark. Then the projector onto states of definite polarization is 
given by 
$
P(a)=(1+\gamma_5\ds{a})/2
$
with the 4-pseudovector $a$,
\beq
{\bf a}=\bzeta+\frac{\bp(\bzeta\cdot\bp)}{m_q(E_k+m_q)}, 
~a^0=\frac{\bzeta\cdot\bp}{m_q}
\label{aa}
\eeq
for a quark moving with the momentum $k=(E_k,\bp)$ \cite{la}. 
The 4-pseudovector
$a$ is reduced into the axial vector $\bzeta$ ($|\bzeta|=1$) 
in the rest frame, which is twice the mean spin vector in the rest
frame. Hence $a$ or $\bzeta$ can specify the polarized state.

The exchange interaction between two quarks with momenta ${\bf k}$ 
and ${\bf q}$ (Fig. ~1) is written as
\beq
f_{{\bf k}\zeta,{\bf q}\zeta'}
=\frac{2}{9}\frac{g^2}{m_q^2}\frac{m_q}{E_k}\cdot \frac{m_q}{E_q}
[2m_q^2-k\cdot q-m_q^2 a\cdot b]\frac{1}{(k-q)^2},
\label{ae}
\eeq
where the 4-pseudovector $b$ is given by the same form as in Eq.~(\ref{aa}) for
the momentum ${\bf q}$. 
The exchange energy is then given by the integration of
the interaction (\ref{ae}) over the two Fermi seas
\footnote{We, here,  don't consider any deformation of Fermi  spheres for
simplicity, while they may be deformed in a realistic case due to the
momentum dependent interaction.}
 for the spin-up
and -down states; eventually, it consists of two contributions,
\beq
\epsilon_{ex}=\epsilon_{ex}^{non-flip}+\epsilon_{ex}^{flip}.
\eeq
The first one arises from the interaction between quarks with the
same polarization, while the second one with the opposite polarization.
The non-flip contribution is the similar one as in electron gas, while 
the flip contribution is a genuine relativistic effect and absent 
in electron gas. We shall see that this relativistic effect
leads to a novel mechanism of ferromagnetism of quark liquid.

\subsection{Symmetry consideration of ferromagnetic phase}

Usual Heisenberg model describes the spin-spin interaction between
adjacent spins localized at lattice points; that is, the Heisenberg
ferromagnet is the spin
alignment in coordinate space. On the other hand, the concept of
spin alignment in quark liquid requires an extension to the phase
space because of the coupling of spin with  momentum. 
Since the spatial part of the quark wave
functions take the plane wave, the
spin orientation is obviously  uniform in coordinate space, once
$\bzeta$ is given. On the other 
hand, the spin does not necessarily take the same orientation in 
momentum space: generally $\bzeta$ should be momentum dependent (see Fig.~2).
\begin{figure}[t]
\centerline{
\epsfysize=4cm
\epsfbox{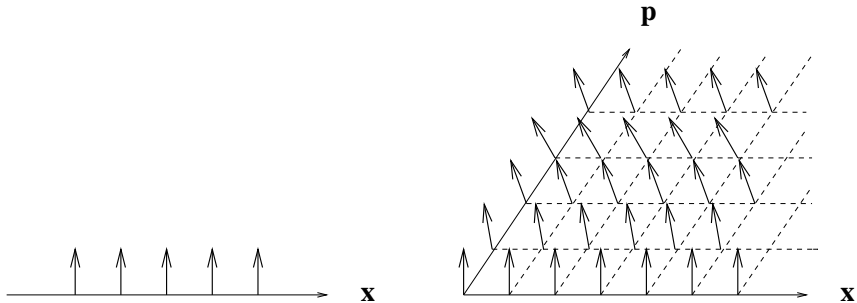}
}
\caption{Heisenberg ferromagnet in the coordinate space 
(left) and quark ferromegnet in the phase space 
(right).}
\end{figure}

The most favorite configuration in momentum space may be determined by 
an energetic consideration, while it seems to be a difficult task. We
consider here only a naive case, where the spin orientation is
uniform even in
the phase space. This is a direct analog of the nonrelativistic version.

Anyway, the ferromagnetic phase is a
spontaneously symmetry broken state with respect to the rotational
symmetry in coordinate space: the order parameter is the mean value of 
$\bzeta$, $\langle\bzeta\rangle$, and symmetry is broken from
$G=O(3)$ to $H=O(2)$ once $\langle\bzeta\rangle$ takes a special 
orientation.

\section{Examples}

We show some results about the total energy of quark liquid, 
$\E_{tot}=\E_{kin}+\E_{ex}$, by adding the kinetic term $\E_{kin}$. Since
gluons have not the flavor quantum numbers, we can consider one flavor 
quark matter without loss of generality. Then quark number density 
directly corresponds to baryon number density, if we assume the three flavor 
symmetric quark matter as mentioned in \S 1.

There are two QCD parameters in our theory: the quark mass $m_q$ and the
quark-gluon coupling constant $\alpha_c$. These values are not well
determined so far. In particular, the concept of quark mass involves subtle
issues; it depends on the current or constituent quark picture and 
may be also related to the existence of chiral phase transition \cite{man}. 
Here we allow some range for these parameters and
take, for example,  a set, $m_q=300$MeV for strange quark 
and $\alpha_c=2.2$, given by the MIT bag model \cite{de}. In Fig.~3 two
results are presented as functions of the polarization parameter $p$
defined by the difference of the number of the spin-up and -down
quarks, $n_q^+-n_q^-\equiv pn_q$. The results clearly show
the first order phase transition, while it is of second order in the
Heisenberg model. The critical
density is around $n_q^c\simeq 0.16$fm$^{-3}$ in this case, 
which corresponds to $n_0$ 
for flavor symmetric quark matter. Note that there is
a metastable ferromagnetic state (the local minimum) 
even above the critical density. 

\begin{figure}[t]
\begin{minipage}{0.45\textwidth}
\epsfysize=4.5cm
\epsfbox{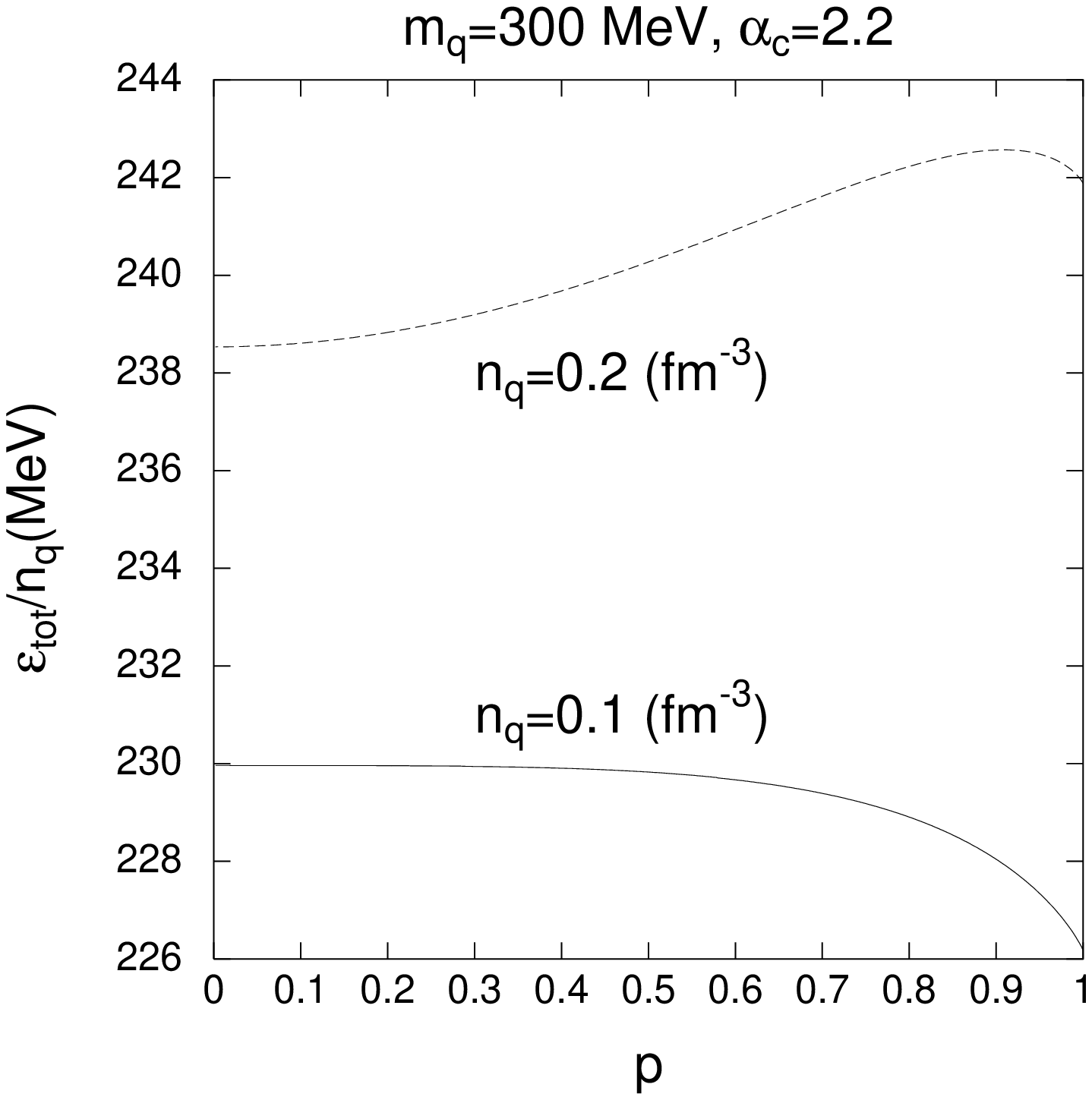}
\caption{Total energy of quark liquid as a function of the
polarization parameter for densities $n_q=0.1, 0.2$fm$^{-3}$.
}
\end{minipage}
\hspace{\columnsep}
\begin{minipage}{0.5\textwidth}
\epsfysize=5cm
\epsfbox{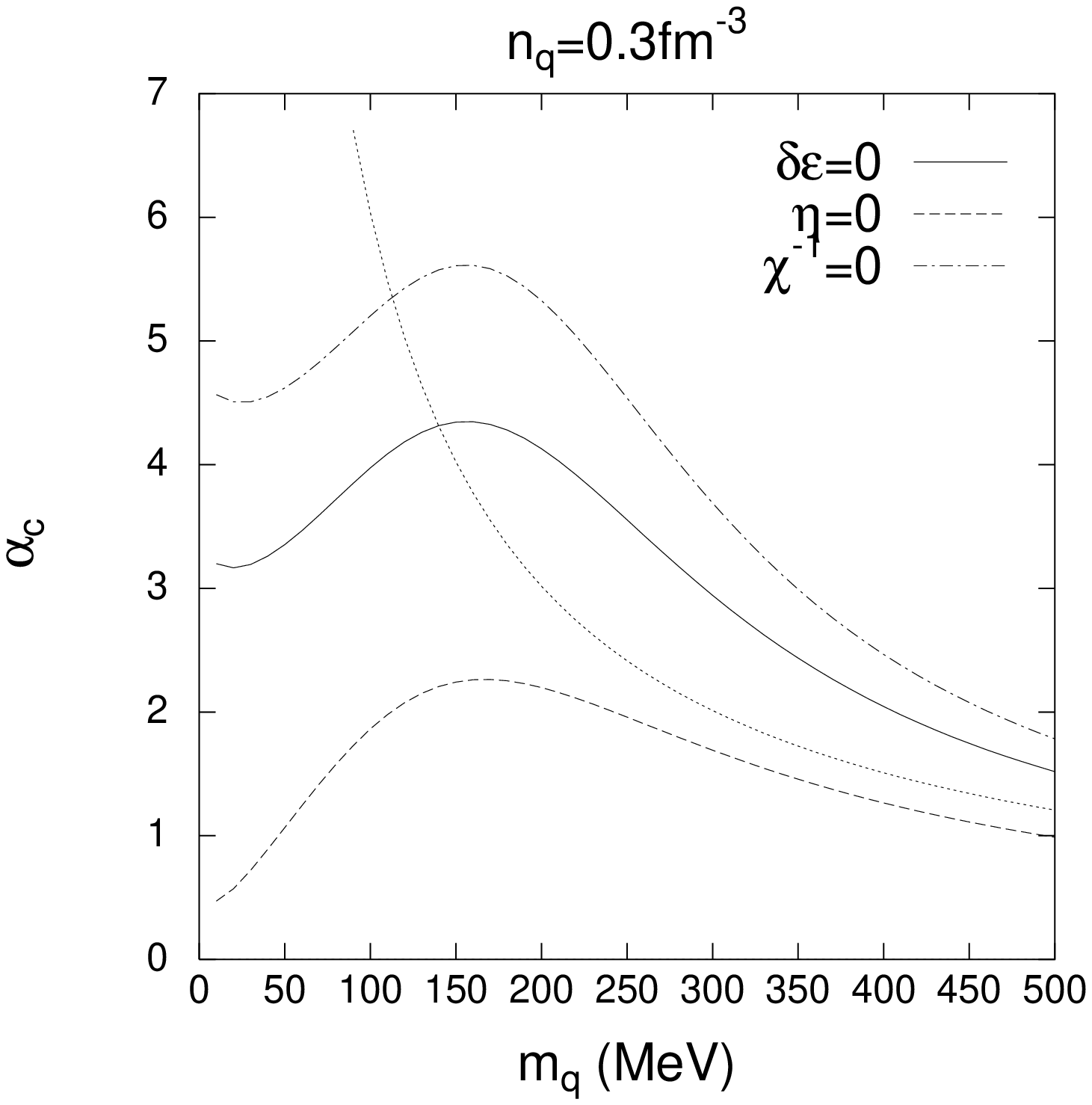}
\vskip 0.3cm
\caption{Phase diagram in the mass ($m_q$)- the coupling constant
($\alpha_c$) plane. $\delta\E$ in the nonrelativistic
calculation is depicted for comparison (the dotted line) .
}
\end{minipage}
\end{figure}
Magnetic properties of quark liquid 
are characterized by three quantities, $\delta \E,
\chi$ and $\eta$;
$\delta \E\equiv \E_{tot}(p=1)-\E_{tot}(p=0)$, which 
is a measure for ferromagnetism to appear in the ground state.
For small $p\ll 1$,
\beq
\E_{tot}-\E_{tot}(p=0)=\chi^{-1} p^2+O(p^4).
\label{ha}
\eeq
$\chi$ is 
proportional to the magnetic susceptibility. In our case it is
less relevant since the phase transition is of first order. Finally, 
$\eta\equiv\partial \E_{tot}/\partial p~|_{p=1}$,
which is a measure for metastability to to exist. 

In Fig.~4 we present a phase diagram in the $m_q - \alpha_c$ plane for
$n_q=0.3$fm$^{-3}$, which corresponds to about twice $n_0$ 
for flavor symmetric quark matter. 
The region above the solid 
line shows the ferromagnetic phase and that bounded by  the
dashed  and dash-dotted lines 
indicates the existence of the matastable state. For 
heavy quarks, which may correspond to the current $s$ quarks
or the constituent quarks before chiral symmetry restoration, 
the ferromagnetic state is favored for small coupling 
constant due to the same mechanism as in electron gas. The
ferromagnetic state is favored again for light quarks, 
which may
correspond to the current $u, d$ quarks, while the nonrelativistic 
calculation never show such tendency. Hence this is due to a 
genuine relativistic effect, where the spin-flip interaction plays an
essential role.

\section{Strange quark star as magnetar}

We have seen that quark matter has a potential to be ferromagnetic at
rather low densities. Here we consider some implications on astrophysics.

Since the idea that nucleons are made of quarks has been confirmed, 
one has expected the existence of quark stars as a third branch of 
compact stars next to the neutron-star branch; 
when pressure or density is increased enough, there should 
occur the deconfinement transition and matter consists of quarks rather than 
nucleons. This naive expectation has been shown to be wrong; 
if the deconfinement transition occurs and quarks
are liberated beyond the maximum central density of neutron stars
(several times of $n_0$), 
they should behave like relativistic and almost free
particles due to the asymptotic freedom of QCD. Thereby the adiabatic
index ($\gamma_{ad}$) of quark matter becomes around $4/3$. 
On the other hand, the
criteria for the gravitationally stable stars reads
$
\gamma_{ad}>4/3+\kappa GM/R, 
$
where the second term means the general relativistic correction, $\sim
0.4$ for $M\simeq 1.4M_\odot$ and $R\simeq 10$Km. 
Hence, the quark-star branch subsequent to
that of neutron stars is impossible. If quark matter exists, it might
occupy only the small portion of the core of neutron stars.

However, there is an alternative idea about quark matter and quark
stars. As first indicated by Chin and Kerman \cite{chi}, 
a large contamination of 
strange quarks are favorable for quark matter at low baryon density
around $0.26$fm$^{-3}$, which is about 1.5 $n_0$.
Their calculation shows that the energy per
baryon of quark matter is larger than that of nucleon, while less than
$\Lambda$ particle.  
Subsequently, Witten and Farhi and Jaffe \cite{far} have pointed out the
possibility that the
almost flavor symmetric quark matter (strange matter) is the ground
state of QCD at finite density within the reasonable range of QCD
parameters. 

Using the idea of strange matter some people suggested that quark
stars with strange matter (strange quark stars) may be possible \cite{mad}. Since the
EOS for strange matter shows the saturation property around 
$n_0$, strange quark stars can have any small radius and mass. 
Thereby, the quark-star branch can be clearly distinguished from that
of neutron stars. 

If a ferromagnetic quark liquid exists stably or metastably around or
above nuclear saturation density, it has
some implications on the properties of strange quark stars and strange
quark nuggets: they should be magnetized in a macroscopic
scale. 
For quark stars with the quark core of $r_q$, simply assuming 
the dipolar magnetic field, we can estimate its strength  at the
surface $R\simeq 10$Km, 
\beq
B_{max}=\frac{8\pi}{3}\left(\frac{r_q}{R}\right)^3\mu_qn_q,
\label{gc}
\eeq
with the quark 
magnetic moment $\mu_q$.
It amounts to order of $O(10^{15-17})$G for $r_q\sim O(R)$ and 
$n_q=O(0.1)$fm$^{-3}$ 
, which should be large enough for magnetars.
A sketch of a strange quark star is presented in Fig~5.

\section{Summary and Concluding remarks}
We have seen that the ferromagnetic phase is realized at low densities
and the metastable state is possible up to rather high densities for a 
reasonable range of the QCD parameters.
\begin{figure}[t]
\centerline{
\epsfysize=4.5cm
\epsfbox{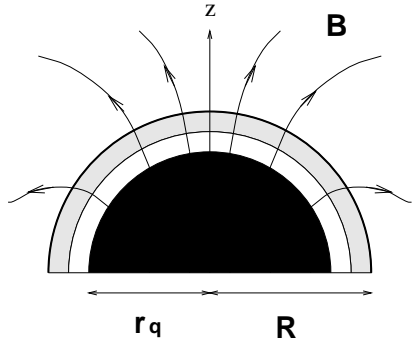}
}
\caption{A model of strange quark star with 
$M\sim 1.4M_\odot$ and $R\sim 10$Km. Almost all the portion is
occupied by strange matter and a small vacuum gap may separate the
quark core from the outer crust, which is composed
of usual solid below the neutron-drip density.}
\end{figure}

We have found that 
ferromagnetic instability is feasible not only
in the massive quark system but also in the light quark system: 
the spin-nonflip contribution is dominant in the nonrelativistic case 
as in electron gas, while a novel mechanism appears  
as a result of the large spin-flip contribution in the relativistic case.

If a ferromagnetic quark liquid exists stably or metastably around or
above nuclear saturation density, strange stars may have a 
strong magnetic field, 
which strength is estimated to be strong enough for magnetars. 
Thereby it might be interesting to model SGR or AXP using our idea. 

Our calculation is basically a perturbative one and the Fermi sea 
remains in a spherical shape. However, if we want to 
get more insight about the
ferromagnetic phase, we must solve the Hartree-Fock equation and
thereby derive a self-consistent mean-field for quark
liquid. Moreover, we need to examine the long range correlation among
quarks by looking into the ring diagrams, which has been known to
be important in the calculation of the susceptibility of electron gas. 

Recently, there have been done many works about the color superconductivity of
quark matter. The order of the energy gap amounts to $O(100)$MeV, while the
energy gain per particle is rather small and several MeV around $n_0$
, which should be the same order of magnitude as that for the 
ferromagnetism \cite{bai}. Hence it may be interesting to explore the phase diagram for
ferromagnetic phase and superconducting phase.


\section*{References}
\begin{thebibliography}{99}
\bibitem{ko}C. Kouveliotou et al., {\it Nature} {\bf 393}, 235~(1998).\\
K. Hurley et al., {\it Astrophys. J.} {\bf 510}, L111~(1999). 

\bibitem{cha} G. Chanmugam, {\it Annu. Rev. Astron. Astrophys.} {\bf
30}, 143~(1992).

\bibitem{pan} V.R. Pandharipande, V.K. Garde and J.K. Srivastava,
{\it Phys. Lett.} {\bf B~38}, 485 ~(1972). 

\bibitem{ta}T. Tatsumi, {\it hep-ph/9910470 (KUNS 1611),
nucl-th/0002014 (KUNS 1636)}.

\bibitem{chi}S.A. Chin and A.K. Kerman, \Journal{\PRL}{43}{1292}{1979}.\\
E. Witten, \Journal{\PRD}{30}{272}{1984}.

\bibitem{far} E. Farhi and R.L. Jaffe, \Journal{\PRD} {30}{2379}{1984}.

\bibitem{mad}for review articles, 
K.S. Cheng, Z.G. Dai and T. Lu, {\it Int. J. Phys.}~{\bf 7}, 139~(1998).\\
J. Madsen, {\it astro-ph/9809032}.


\bibitem{bl}F. Bloch, {\it Z. Phys.} ~{\bf 57},~545 ~(1929).

\bibitem{la} V.B. Berestetsii, E.M. Lifshitz and L.P. Pitaevsii,\\ 
{\it Relativistic Quantum Theory}(Pergamon Press, 1971).

\bibitem{man} A. Manohar and H. Georgi, {\it Nucl. Phys.}{\bf B 234},
189~(1984).

\bibitem{de}T. DeGrand et al., \Journal{\PRD}{12}{2060}{1975}.

\bibitem{bai}D. Bailin and A. Love, {\it Phys. Rep.}~{\bf 107},
325~(1984).\\
J. Berges and K. Rajagopal, {\it Nucl. Phys.}~{\bf B 538}, 215~(1999).
\end{thebibliography}
\end{document}